\documentclass[twocolumn]{aastex63}

\usepackage{color}
\usepackage{graphicx,graphics}
\usepackage{amssymb,amsmath}
\usepackage{natbib}
\usepackage {threeparttable}
\usepackage {bm}

\newcommand{\Mstar}{M_\star}

\shorttitle{Catch Me if You Can: Biased Distribution of
Ly$\alpha$-emitting Ealaxies}
\shortauthors{R. Momose et al.}

\begin{document}

\title{Catch Me if You Can: Biased Distribution of
Ly$\alpha$-emitting Galaxies according to the Viewing Direction}


\correspondingauthor{Rieko Momose}
\email{momose@astron.s.u-tokyo.ac.jp}

\author[0000-0002-8857-2905]{Rieko Momose}
\affiliation{Department of Astronomy, School of Science, The University of Tokyo, 
7-3-1 Hongo, Bunkyo-ku, Tokyo 113-0033, Japan}

\author[0000-0002-2597-2231]{Kazuhiro Shimasaku}
\affiliation{Department of Astronomy, School of Science, The University of Tokyo, 
7-3-1 Hongo, Bunkyo-ku, Tokyo 113-0033, Japan}
\affiliation{Research Center for the Early Universe, The University of Tokyo, 7-3-1 Hongo, Bunkyo-ku, Tokyo 113-0033, Japan}

\author[0000-0001-7457-8487]{Kentaro Nagamine}
\affiliation{Department of Earth and Space Science, Osaka University, 1-1 Machikaneyama, Toyonaka, Osaka 560-0043, Japan}
\affiliation{Department of Physics and Astronomy, University of Nevada, Las Vegas, NV 89154-4002, USA}
\affiliation{Kavli-IPMU (WPI), The University of Tokyo, 5-1-5 Kashiwanoha, Kashiwa, Chiba 277-8583, Japan}

\author{Ikkoh Shimizu}
\affiliation{Shikoku Gakuin University, 3-2-1 Bunkyocho, Zentsuji, Kagawa 765-0013, Japan}

\author[0000-0003-3954-4219]{Nobunari Kashikawa}
\affiliation{Department of Astronomy, School of Science, The University of Tokyo, 
7-3-1 Hongo, Bunkyo-ku, Tokyo 113-0033, Japan}
\affiliation{Research Center for the Early Universe, The University of Tokyo, 7-3-1 Hongo, Bunkyo-ku, Tokyo 113-0033, Japan}

\author{Makoto Ando}
\affiliation{Department of Astronomy, School of Science, The University of Tokyo, 7-3-1 Hongo, Bunkyo-ku, Tokyo 113-0033, Japan}

\author[0000-0002-3801-434X]{Haruka Kusakabe}
\affiliation{Observatoire de Gen\`{e}ve, Universit\'e de Gen\`{e}ve, 51 chemin de P\'egase, 1290 Versoix, Switzerland}


\begin{abstract}

We report that Ly$\alpha$-emitting galaxies (LAEs) 
may not faithfully trace the cosmic web of neutral hydrogen ({\sc Hi}), but their distribution is likely biased depending on the viewing direction.
We calculate the cross-correlation (CCF) between galaxies and Ly$\alpha$ forest transmission fluctuations on the near and far sides of the galaxies separately, for three galaxy samples at $z\sim2$: LAEs, [{\sc Oiii}] emitters (O3Es), and continuum-selected galaxies.
We find that only LAEs have anisotropic CCFs, with the near side one showing lower signals up to $r=3-4~h^{-1}$~comoving Mpc. 
This means that the average {\sc Hi} density on the near side of LAEs is lower than that on the far-side by a factor of $2.1$ under the Fluctuating Gunn-Peterson Approximation.
Mock LAEs created by assigning Ly$\alpha$ equivalent width ($EW_\text{Ly$\alpha$}^\text{obs}$)
values to O3Es with an empirical relation also show similar, anisotropic CCFs if we use only objects with higher
$EW_\text{Ly$\alpha$}^\text{obs}$
than a certain threshold.
These results indicate that galaxies on the far side of a dense region are more difficult to be detected (``hidden'') 
in Ly$\alpha$
because Ly$\alpha$ emission toward us is absorbed by dense neutral hydrogen.
If the same region is viewed from a different direction, a different set of LAEs will be selected as if galaxies are playing hide-and-seek using {\sc Hi} gas.
Care is needed when using LAEs to search for overdensities.
\end{abstract}

\keywords{galaxies: formation -- evolution -- intergalactic medium, quasars: absorption lines, cosmology: large-scale structure of universe}

\section{Introduction}

Matter in the universe is not uniformly distributed in space but forms large-scale filamentary structure or the cosmic web. Many galaxy redshift surveys have depicted the cosmic web of the present-day universe (e.g., \citealp{deLapparent86,Tegmark04,Guzzo14,Libeskind18}). 
For the young universe, the most successfully used galaxies to map the web are those with hydrogen Ly$\alpha$ emission at $1216$ {\AA} (Ly$\alpha$-emitting galaxies, LAEs).
LAEs are one of the major galaxy populations in the young universe (e.g., \citealp{hu98,Malhotra04,Shibuya18}). LAEs in the post reionization epoch ($z\leq6$) can be easily detected by narrow-band imaging whose wavelength is tuned to Ly$\alpha$ emission at the target redshift. 
Many LAE surveys have revealed filamentary structures and proto-clusters in the young universe (e.g., \citealp{Shimasaku03,Shimasaku04,cai17}
).

However, several studies have recently pointed out that the density peak of LAEs' distribution does not always match that traced by other galaxy populations. For instance, \citet{shimakawa17b} have found a few Mpc deviation of LAEs' density peak in a $z=2$ proto-cluster from that of H$\alpha$-emitting galaxies, which are normal star-forming galaxies commonly seen at high redshifts.
A similar offset from continuum-selected galaxies has also been reported \citep{Overzier08,toshikawa16,shi19}.
More importantly, \citet{momose20b} have suggested that the peak of LAEs' distribution does not match that of intergalactic neutral hydrogen (IGM {\sc Hi}). 
With its density increasing with matter density, IGM {\sc Hi} is a faithful tracer of the cosmic web that has recently come into use, although the areas mapped by IGM {\sc Hi} are still limited because the observations are very time-consuming. These reported discrepancies might imply that LAEs intrinsically avoid dense parts of the cosmic web for some reason (e.g., \citealp{shimakawa17b,oteo18,shi19,momose20b}). 

Alternatively, the discrepancies may be a bias caused by the absorption of Ly$\alpha$ photons by IGM {\sc Hi} in dense regions (e.g., \citealp{Dijkstra07,Zheng11a,Laursen11,GL20a,hayes20}). In this case, a biased distribution of LAEs along the line-of-sight is expected.
LAEs located on the far side of a dense region are more difficult to detect because their Ly$\alpha$ emission toward us is heavily absorbed by IGM {\sc Hi} during passing through the region \citep{Zheng11a,GL20a}. As a result, fewer LAEs will be detected on the far side even if the original number is identical. So far, however, no observation has reported such an anisotropic distribution.

Aiming at finding such anisotropic features, we investigate the connection between LAEs and IGM {\sc Hi} with a publicly available 3D IGM tomography map of the Ly$\alpha$ forest transmission fluctuation ($\delta_\text{F}$), called the COSMOS Ly$\alpha$ Mapping And Tomography Observations (CLAMATO, \citealp{lee16,lee18}), as an extension of \citet{momose20b}. 
This paper is organized as follows. 
We introduce the data and methodology used in this study in Section~\ref{sec:data_ana} and present the results in Section~\ref{sec:result}. 
An examination of CCFs using mock LAEs is presented in Section~\ref{sec:dis_mocklae}. We summarize our study together with discussions in Section~\ref{sec:summ}.
We use a cosmological parameter set of 
($\Omega_\text{m}$, $\Omega_\text{$\Lambda$}$, $h$) = ($0.31$, $0.69$, $0.7$)
throughout this paper \citep{lee16,lee18}.
All distances are described in a comoving unit unless otherwise stated. We indicate ``cosmic web'' and ``IGM'' as those traced by {\sc Hi} unless otherwise specified.



\begin{figure*}
	\begin{center}
	\includegraphics[width=\linewidth]{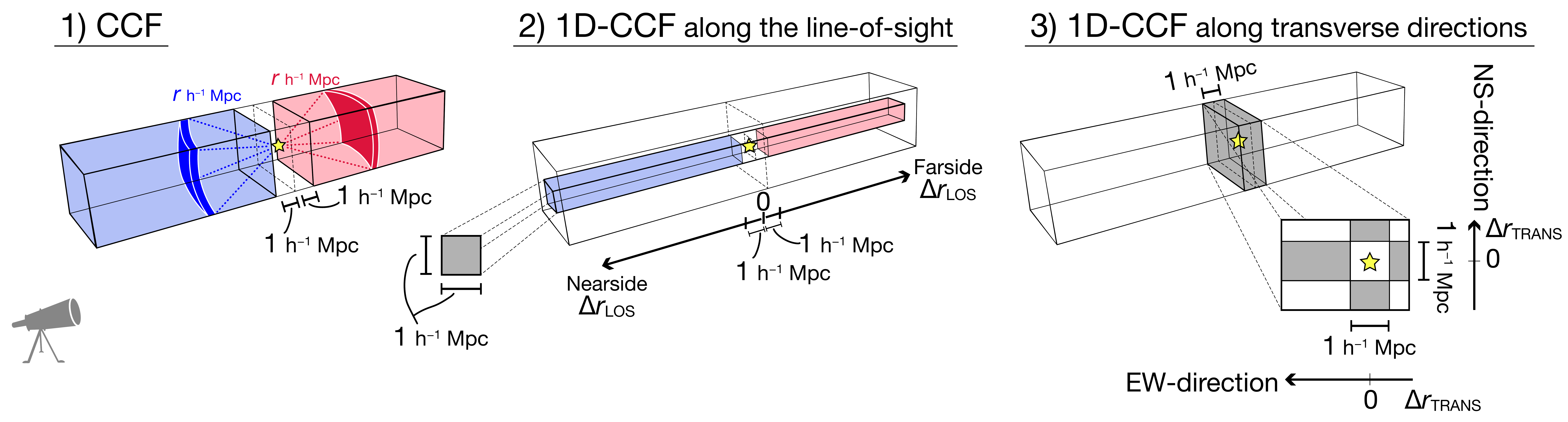} 
	\caption{
    Schematic diagrams of the CLAMATO data used in each analysis. The CCF and the 1D-CCFs along the line-of-sight and transverse directions are displayed from left to right. Note that the CLAMATO spans comoving dimensions of ($x, y, z$) $=$ ($30, 24, 438$)~$h^{-1}$~cMpc. The sub-volume of the CLAMATO used in each analysis is colored in light blue (near side), faint red (far side), and gray (two transverse directions).
    The exact area used to calculate the CCF at radius $r$ is colored in dark blue and dark red for the near and far sides, respectively.
	}
	\label{fig:analysis}
	\end{center}
\end{figure*}


\section{Data and Analyses}
\label{sec:data_ana}

\subsection{IGM {\sc Hi} Data}
\label{sec:igmdata}

We use the  CLAMATO\footnote{The data is from: http://clamato.lbl.gov} as a tracer of IGM {\sc Hi} \citep{lee16,lee18} .
The CLAMATO is a 3D data-cube of $\delta_\text{F}$ over $z=2.05-2.55$, which is reconstructed from $240$ galaxies and quasars spectra taken by the LRIS on the Keck I \citep{oke95,steidel04}. Here, $\delta_\text{F}$ is defined by 
\begin{equation}
    \delta_\text{F}=\frac{F}{\langle F_z \rangle} - 1,
    \label{eq:delF}
\end{equation}
where $F$ and $\langle F_z\rangle$ are the Ly$\alpha$ forest transmission and its cosmic mean from \citet{FG08}. Hence, $\delta_\text{F}$ is the excess transmission to Ly$\alpha$ photons. Higher IGM {\sc Hi} densities give lower $F$, and negative (positive) $\delta_\text{F}$($x$) values indicate higher (lower) {\sc Hi} densities than the cosmic mean. 
The effective transverse separation of the CLAMATO is $2.04$ comoving~$h^{-1}$~Mpc (hereafter $h^{-1}$~cMpc). 
The spatial resolution in the line-of-sight direction is $2.35~h^{-1}$~cMpc at $z=2.3$. The final 3D data cube of the CLAMATO spans ($x$, $y$, $z$) $=$ ($30$, $24$, $438$)~$h^{-1}$~cMpc with $0.5~h^{-1}$~cMpc pixel size.

\subsection{Galaxy Samples}
\label{sec:galcatalog}

We use three galaxy samples constructed in \citet{momose20b}.
One is a sample of 19 LAEs whose redshifts have been determined by nebular lines of H$\alpha$ and/or [{\sc Oiii}]\footnote{Although some of the LAEs are also bright in [{\sc Oiii}] emission, we refer to them as LAEs based on their first identification from narrow-band images.} \citep{naka12,nakajima13,hashimoto13,shibuya14b,konno16}. 
The other two used as control samples are $85$ [{\sc Oiii}]$\lambda\lambda 4959$, $5007$ emitting galaxies (O3Es), and $570$ continuum-selected galaxies with a spectroscopic redshift.
O3Es and continuum-selected galaxies are also star-forming galaxies like LAEs, but their selections, based on [{\sc Oiii}] and far-ultraviolet luminosities, respectively, are not affected by IGM {\sc Hi}.

 
\subsection{Analyses}
\label{sec:analysis}

We conduct two different analyses. 
The first is the cross-correlation (CCF) between galaxies and the CLAMATO using the same definition of our previous work \citep{momose20b}: 
\begin{equation}
\begin{split}
  \xi_\text{$\delta$F}(r) 
  &= \frac{1}{\sum_{i=1}^{N(r)} \omega_{F, i}} \sum_{i=1}^{N(r)} \omega_{F, i} \delta_{F, i} \\
  &- \frac{1}{\sum_{j=1}^{M(r)} \omega_{ran, j}} \sum_{j=1}^{M(r)} \omega_{ran, j} \delta_{ran, j},
  \label{eq:cross-corr}
\end{split}
\end{equation}
\begin{equation}
    \omega_{F, i} = \frac{1}{(\sigma_{F, i})^2}, \   
    \omega_{ran, j} = \frac{1}{(\sigma_{ran, j})^2}, 
    \label{eq:cross-corr-sigma}
\end{equation}
where $\xi_\text{$\delta$F}$ is the cross-correlation at a separation $r$; 
$\delta_{F, i}$ and $\delta_{ran, j}$ are the transmission fluctuation at places $i$ and $j$ separated by $r$ from a galaxy and random point, respectively, in question with 
$\sigma_{F, i}$ and $\sigma_{ran, j}$ being their errors;
$N(r)$ and $M(r)$ represent the numbers of pixel-galaxy and pixel-random pairs with separation $r$, respectively. 
The $\sigma_{F}$ and $\sigma_{ran}$ are evaluated with the CLAMATO's 3D noise standard deviation measurements, including pixel noise, finite skewer sampling, and the intrinsic variance of the Ly$\alpha$ forest \citep{lee18}.
Unlike our previous studies \citep{momose20b}, we calculate the CCFs on the near and far sides of galaxies separately. The CCF on the near (far) side is derived only using the data in the near (far) hemispheres (regions colored in blue and red of Figure~\ref{fig:analysis}--1). For both calculations, the data within a line-of-sight separation of $1~h^{-1}$~cMpc from each galaxy is excluded to eliminate the influence of {\sc Hi} in the circumgalactic medium (CGM).

The second analysis is to 
compute a one dimensional CCF (hereafter ``1D-CCF'') defined by
\begin{equation}
  \langle\delta_\text{F}\rangle(\Delta r) = \frac{1}{\sum_{i=1}^{N(\Delta r)} \omega_{F, i}} \sum_{i=1}^{N(\Delta r)} \omega_{F, i} \delta_{F, i},
    \label{eq:weighted_mean}
\end{equation}
for the line-of-sight and two transverse directions.
For the line-of-sight direction, we use narrow tomography data of the projected $1\times1$ ($h^{-1}$~cMpc)$^2$ area centered at individual galaxies (Figure~\ref{fig:analysis}--2).
Note that regions colored in light blue and red are referred to as the near and far sides.
Similarly, for the transverse directions, we use narrow data of a cross-section of $1\times1$ ($h^{-1}$~cMpc)$^2$ along with the North-South (NS) and East-West (EW) directions (Figure~\ref{fig:analysis}--3).
We measure $\langle\delta_\text{F}\rangle$($\Delta r$) in $2.5~h^{-1}$~cMpc steps over $-16\leq(\Delta r/h^{-1}$~cMpc)~$\leq16$ for the line-of-sight direction and $-11\leq(\Delta r/h^{-1}$~cMpc)~$\leq11$ for the transverse directions, excluding $-1\leq(\Delta r/h^{-1}$~cMpc)~$\leq1$.

In both analyses, errors are estimated with Jackknife resampling by removing one object. Therefore, the number of Jackknife samples is the same as that of galaxies in the original sample. 
Although Jackknife resampling for the cross-correlation is usually evaluated by dividing the survey volume into several small subvolumes, to be conservative, we do not use this resampling in this study because it gives smaller errors than obtained above \citep{momose20b}.


\begin{figure*}
	\begin{center}
	\includegraphics[width=\linewidth]{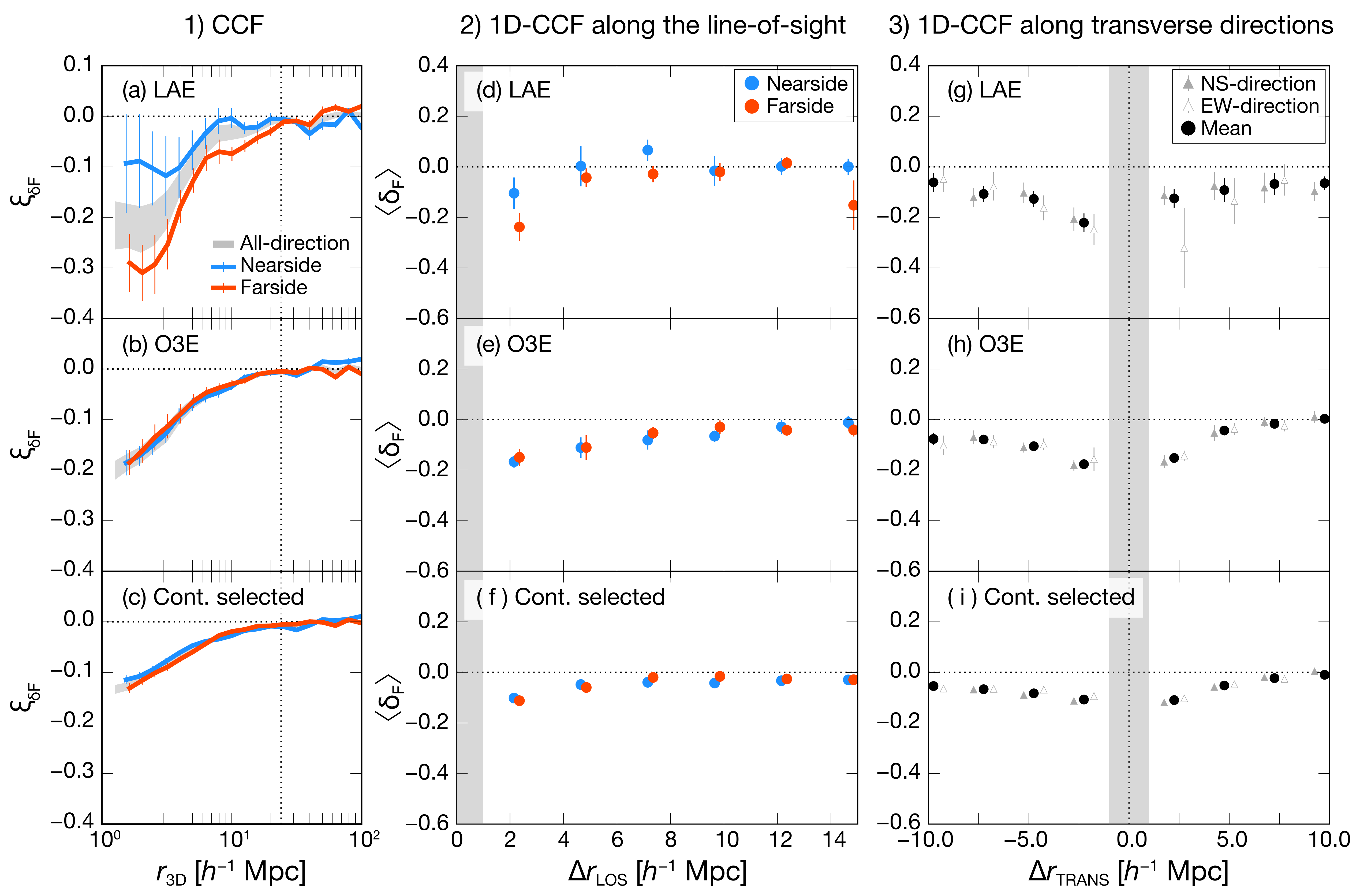} 
	\caption{CCF (panels (a)--(c)) and 1D-CCFs along the line-of-sight (panels (d)--(f)) and
	two transverse directions (panels (g)--(i)) for LAEs, O3Es, and continuum-selected galaxies.
    \textit{Panels (a)--(c)}. Blue and red lines indicate, respectively, the CCFs on the near and far sides of galaxies. Gray shades represent the CCFs calculated using all directions \citep{momose20b}. A vertical dotted line indicates the possible largest radius for 3D cross-correlation calculations.
    \textit{Panels (d)--(f)}. $\langle\delta_\text{F}\rangle$ along the line-of-sight direction in $2.5$~$h^{-1}$~cMpc steps over $1\leq$ ($\Delta r$/$h^{-1}$~cMpc) $\leq16$, plotted against the absolute distance from galaxies $\Delta r_\text{LOS}$, with blue (red) color indicating the near (far) side of them. A gray zone indicates the distance range excluded from the analysis.  
    \textit{Panels (g)--(i)}. $\langle\delta_\text{F}\rangle$ along two transverse directions in $2.5$~$h^{-1}$~cMpc steps over $-11\leq$ ($\Delta r$/$h^{-1}$~cMpc) $\leq11$, except the range in $-1\leq$ ($\Delta r$/$h^{-1}$~cMpc) $\leq1$, where $\Delta r_\text{TRANS}$ indicates the distance from galaxies shown in Figure~\ref{fig:analysis}-3.
    Filled and open triangles denote $\langle\delta_\text{F}\rangle$ of the NS- and EW-directions, respectively, while filled circles indicate the average of the two directions. 
    }
	\label{fig:result_1}
	\end{center}
\end{figure*}


\section{Results}
\label{sec:result}


The results of the two analyses are shown in Figure~\ref{fig:result_1}. 
Figure~\ref{fig:result_1} (a) shows that the near side CCF of LAEs is consistently higher than the far side one up to $r=3-4~h^{-1}$~cMpc, possibly up to $r\sim20$~$h^{-1}$~cMpc, meaning that the average density of IGM {\sc Hi} gas around LAEs is systematically lower on the near side. 
The IGM {\sc Hi} density at $3~h^{-1}$~cMpc between the two sides differs by a factor of $2.1$ under the Fluctuating Gunn--Peterson Approximation (Appendix~\ref{sec:app_FGPA}).
In contrast, no such systematic difference is seen in the other two galaxy populations (Figure~\ref{fig:result_1} (b) and (c)).

A similar trend is found in the 1D-CCF along the line-of-sight in Figure~\ref{fig:result_1}~(d)--(f). 
Although the difference is marginal, the 1D-CCF of LAEs is higher on the near side up to $\sim8~h^{-1}$~cMpc.
In contrast, the 1D-CCFs of O3Es and continuum-selected galaxies are nearly symmetric around $\Delta r=0$, indicating an isotropic {\sc Hi} density distribution along the line-of-sight with respect to the position of those galaxies. 
Unlike the 1D-CCF along the line-of-sight, those along transverse directions  (Figure~\ref{fig:result_1}~(g)--(i)) show nearly symmetric distributions within the $1\sigma$ errors with a negative peak at $r=0$, indicating an isotropic {\sc Hi} density distribution.

We calculate the probability of an anisotropic CCF and 1D-CCF along the line-of-sight like those seen in LAEs being caused by chance due to a small sample, by conducting the same analyses to randomly selected $19$ continuum-selected galaxies and O3Es. The probability is found to be only $2\%$ even if we loosen the criteria for anisotropy.\footnote{We set two criteria. The first is that the near side CCF is higher than the far side one beyond the error bars at any radius up to $r=4~h^{-1}$~cMpc. The second criterion is the same as the first one except for the innermost radius by considering the case in mock LAEs presented in Section~\ref{sec:results_mockLAE}. The probability of satisfying each criterion is $0$ and $2\%$ ($1$ and $2\%$) for continuum-selected galaxies (O3Es).}
Therefore, the anisotropy of the {\sc Hi} density found for LAEs is unlikely to be by chance.

One may be concerned that the limited spatial distribution of our LAEs along the transverse directions (($\Delta x$, $\Delta y$) = ($30$, $24$)$~h^{-1}$~cMpc) and anlog the line-of-sight ($z=2.14-2.22$\footnote{This redshift range is defined by the full-width-at-half-maximum (FWHM) of the narrow-band filter used for our LAE search \citep{naka12,konno16}.
}
or $69.7~h^{-1}$~cMpc) causes an anisotropic {\sc Hi} density distribution along the line-of-sight.
We also conduct the same analyses to $17$ O3Es and $108$ continuum-selected galaxies in this redshift range and find anisotropy in the CCF and 1D-CCF along the line-of-sight around O3Es, but with an opposite sign to that of LAEs: the {\sc Hi} density is lower on the far side than on the near side. 
Besides, the probability that randomly selected galaxies reproduce an anisotropic CCF and 1D-CCF along the line-of-sight like those found for real LAEs is less than $1\%$. These results indicate that the anisotropy seen around LAEs is unlikely to be due to a small sample in a limited redshift range.


%
\section{A toy model with mock Ly$\alpha$-emitting galaxies}
\label{sec:dis_mocklae}
%


\begin{figure} 
	\begin{center}
	\includegraphics[width=\linewidth]{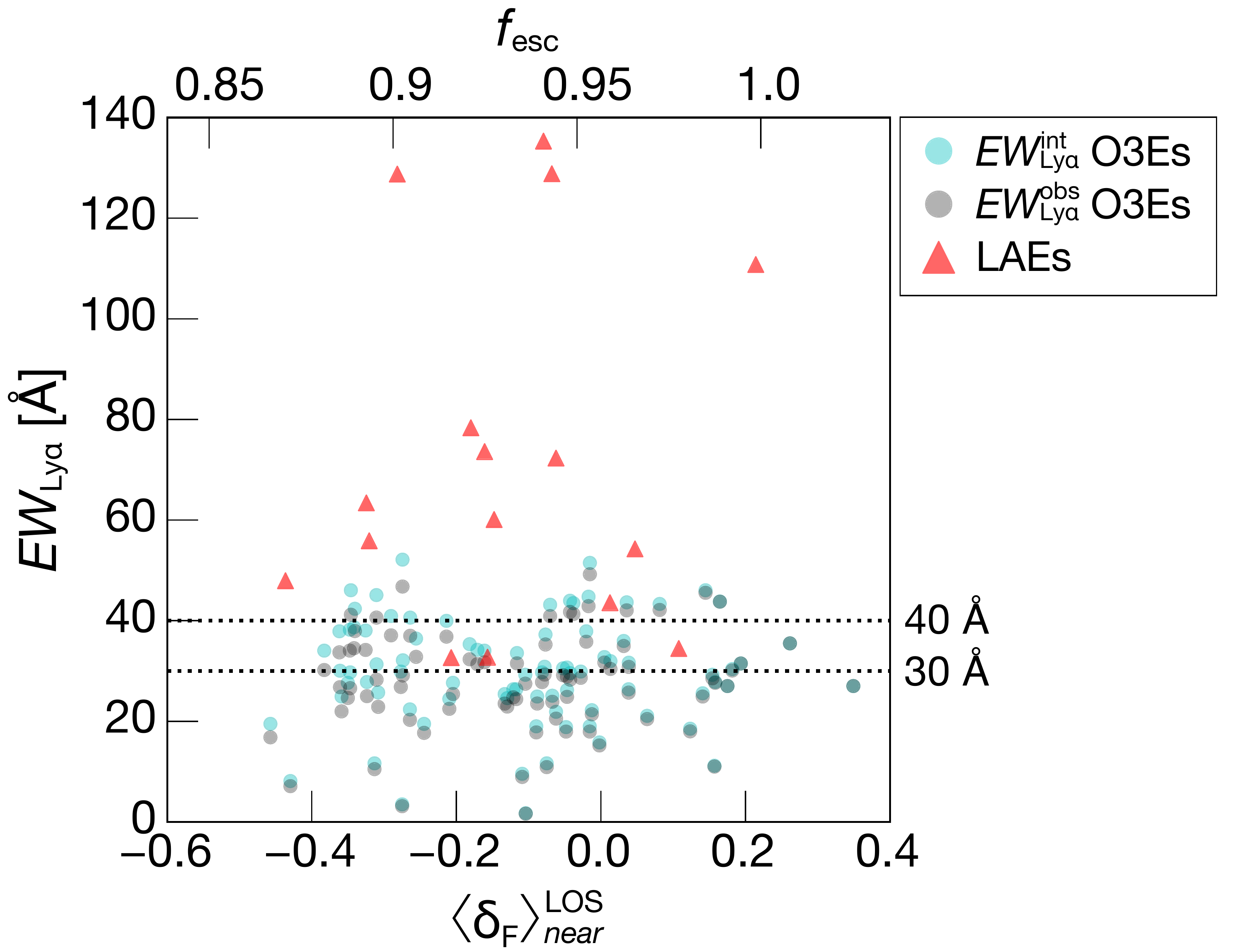} 
    \caption{$EW_\text{Ly$\alpha$}$ as a function of $\langle\delta_\text{F}\rangle$ on the near side over 
    $-5.0\leq(\Delta r/h^{-1}$~{\rm cMpc})~$\leq-1.7$
    along the line-of-sight direction which we refer to as 
    $\langle\delta_\text{F}\rangle^\text{LOS}_{near}$.
	Cyan and black circles represent, respectively,  $EW_\text{Ly$\alpha$}^\text{int}$ and $EW_\text{Ly$\alpha$}^\text{obs}$ of O3Es used to create mock LAEs. 
    The observed $EW_\text{Ly$\alpha$}$ of our LAEs are also plotted as filled triangles except for three objects with $EW_\text{Ly$\alpha$}^\text{obs}>140$ {\AA}.
	Dashed lines show the two thresholds for selecting mock LAEs. 
	The escape fraction of Ly$\alpha$ photons through the IGM is calculated as $f_\text{esc} = EW_\text{Ly$\alpha$}^\text{obs} / EW_\text{Ly$\alpha$}^\text{int}$.
    }
	\label{fig:dist_EW}
	\end{center}
\end{figure}

\begin{figure*}
	\begin{center}
	\includegraphics[width=\linewidth]{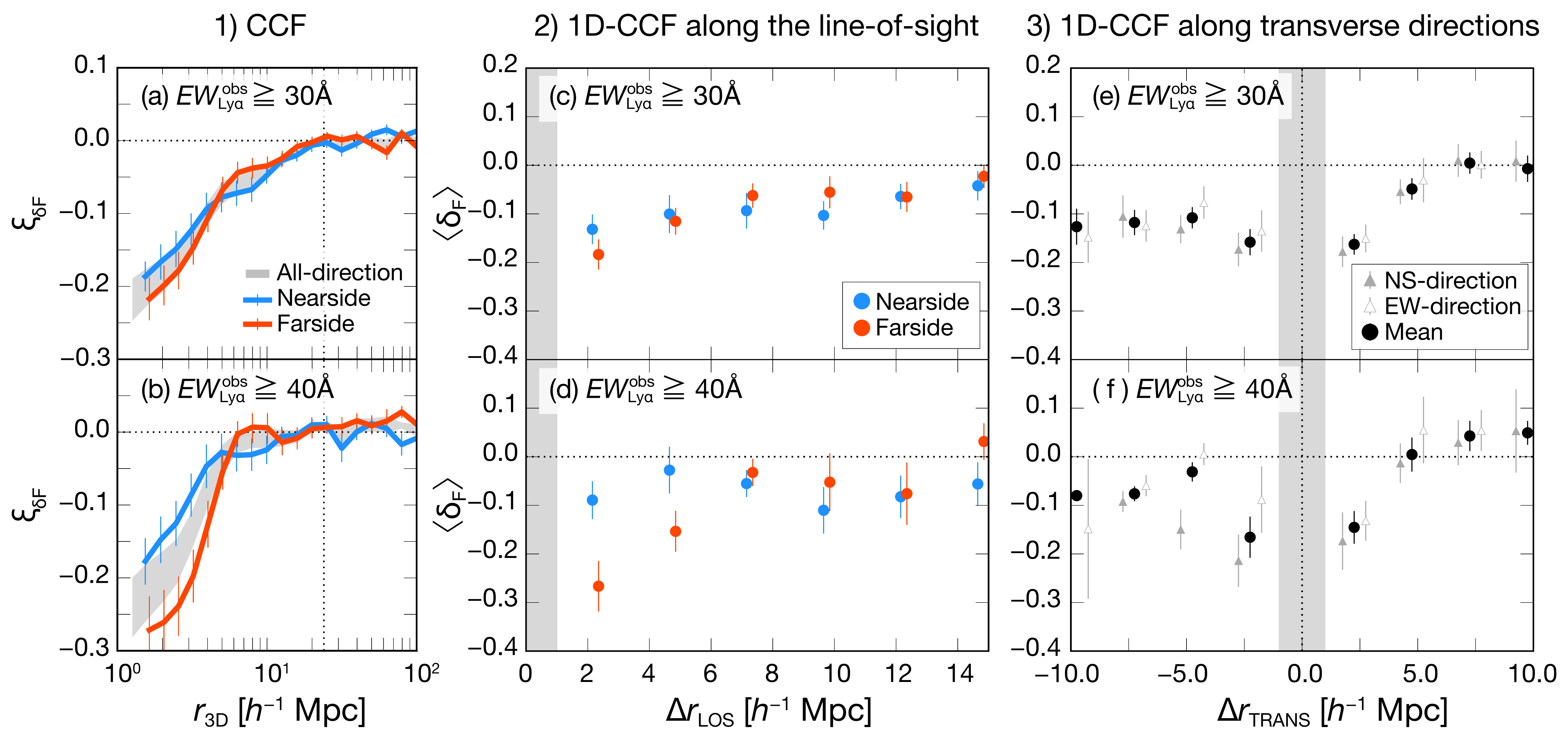} 
	\caption{Same as Figure~\ref{fig:result_1}, but for mock LAEs, which are O3Es with $EW_\text{Ly$\alpha$}^\text{obs}$ greater than $30$ or 
	$40$ {\AA}. 
	}
	\label{fig:IM_mock}
	\end{center}
\end{figure*}


We show in Section~\ref{sec:result} that the IGM {\sc Hi} density distribution around LAEs is anisotropic, with the density on the near side being lower than on the far side. 
Considering that the {\sc Hi} density averaged over all directions decreases with the distance from LAEs, this result leads to the picture that LAEs for a given dense region are preferentially distributed on its near side; at any position on the near side, the {\sc Hi} gas in front of that position should be more transparent to Ly$\alpha$ than on the opposite side. 

The most natural explanation of this anisotropy is a selection effect. LAEs on the far side of dense regions are more difficult to detect because Ly$\alpha$ photons emitted from them toward us have to pass through the dense regions of low Ly$\alpha$ transmission. 
To test this hypothesis, we construct a toy model with mock LAEs and examine if galaxies, whose intrinsic distribution is normal, show an anisotropic {\sc Hi} density distribution like actual LAEs if the sample is limited to those with strong Ly$\alpha$ emission in the observed-frame owing to their location relative to a dense region.

\subsection{Methodology}
\label{sec:methodology_mockLAE}

Our toy model is based on a mock LAE sample created from the O3Es.
We carry out the examination as follows.
First, we assign each object of the O3E sample an ``intrinsic'' equivalent width, $EW_\text{Ly$\alpha$}^\text{int}$ (hence an ``intrinsic'' Ly$\alpha$ luminosity) according to an empirical $EW_\text{Ly$\alpha$}$--stellar mass ($\Mstar$) relation by  \citet{Cullen20}: 
\begin{equation}
    \label{eq:cullen20}
    EW_\text{Ly$\alpha$}^\text{int} = - \frac{0.24 \log{\Mstar} - 4.65}{0.015} .
\end{equation}
We use $\Mstar$ estimates in the photometric redshift catalog of \cite{straa16}.
Note that we adopt not the best-fit relation of \citet{Cullen20} but its upper $1\sigma$ envelope because the best-fit relation assigns a large fraction of the sample negative $EW_\text{Ly$\alpha$}^\text{int}$,\footnote{We also note that although we regard $EW_\text{Ly$\alpha$}$ in the relation as an ``intrinsic equivalent width ($EW_\text{Ly$\alpha$}^\text{int}$)'' after absorption by the interstellar medium (ISM) and the CGM, it may also be affected by the IGM attenuation.
}
which cannot be used for selecting mock LAEs. Nonetheless, this prescription is consistent with the fact that the Ly$\alpha$ luminosities of LAEs are boosted compared with other star-forming galaxies.

Next, we attenuate the Ly$\alpha$ luminosity 
of each object 
using the actual line-of-sight $\delta_\text{F}$ at the position of this object to obtain an ``observed'' equivalent width ($EW_\text{Ly$\alpha$}^\text{obs}$) by taking the attenuation by the IGM into account:
\begin{equation}
    EW_\text{Ly$\alpha$}^\text{obs} = 0.24 \  EW_\text{Ly$\alpha$}^\text{int} \  \langle F \rangle_{near}^\text{LOS} + 0.76 \  EW_\text{Ly$\alpha$}^\text{int} ,
\label{eq:ewobs}
\end{equation}
where 
$\langle F \rangle_{near}^\text{LOS}$ 
is the transmission in the line-of-sight direction averaged over 
$-5.0\leq(\Delta r/h^{-1}$~cMpc) $\leq-1.7$
on the near side, in a projected square of $1\times1$ ($h^{-1}$~cMpc)$^2$ area center at the galaxy.
Equation~\ref{eq:ewobs} comes from the following consideration.
First, we assume a double-peaked Ly$\alpha$ profile where the red peak is located $200$ km s$^{-1}$ redward of $\lambda$(Ly$\alpha$) while the blue peak, contributing $24\%$ of the total Ly$\alpha$ emission, is centered at $-200$ km s$^{-1}$ with a $200$ km s$^{-1}$ width.
This profile is based on a stacked Ly$\alpha$ spectrum of $z=2.2$ LAEs presented in \citet{Matthee21} but also roughly consistent with other observations of individual LAEs (e.g., \citealp{rakic11,hashimoto13,hashimoto15}).
We then assume that only the blue peak is subject to the IGM attenuation, neglecting the effect of the peculiar motion of galaxies relative to the IGM. 
Because Ly$\alpha$ photons in the blue peak, which have relative velocities of $-300$ to $-100$~km~s$^{-1}$, are redshifted to $\lambda$(Ly$\alpha$) after traveling over $1.7 \leq \Delta r/h^{-1} {\rm cMpc} \leq 5.1$, we use the line-of-sight transmission averaged over 
this distance range, $\langle F \rangle_{near}^\text{LOS}$. 
We calculate $\langle F \rangle_{near}^\text{LOS}$ from the excess transmission averaged over the same distance range, $\langle\delta_\text{F}\rangle_{near}^\text{LOS}$, by:
\begin{equation}
    \langle F \rangle_{near}^\text{LOS} = ( 1 + \langle\delta_\text{F}\rangle_{near}^\text{LOS} ) \langle F_z \rangle.
\end{equation}
We assume 
$\langle F \rangle_{near}^\text{LOS}=1$
for four O3Es whose 
$\langle F \rangle_{near}^\text{LOS}$
is larger than unity so that their $EW_\text{Ly$\alpha$}^\text{obs}$ is equal to $EW_\text{Ly$\alpha$}^\text{int}$.
We should also note that an observed Ly$\alpha$ luminosity is determined not only by the line-of-sight IGM {\sc Hi} absorption but to some degree by the absorption in all other directions, as pointed out by \citet{Zheng11a}.
However, since it is extremely complicated to evaluate the impact of Ly$\alpha$ suppression in all directions, we consider the line-of-sight direction alone.

The resultant $EW_\text{Ly$\alpha$}^\text{obs}$ estimates, presented by black circles in Figure~\ref{fig:dist_EW}, are lower than those of actual LAEs plotted as red stars.
It is probably because LAEs generally have higher Ly$\alpha$ ISM/CGM escape fractions than continuum and nebular emission line selected galaxies, owing to their lower dust extinction (e.g., \citealp{Stark10,Wardlow14,Kusakabe15}), lower {\sc Hi} column density of the ISM (e.g., \citealp{hashimoto15}), and/or higher ionization parameter (e.g., \citealp{Nakajima14,sobral18a}).
However, our goal here is not to accurately reproduce observed LAEs but to create mock galaxies whose $EW_\text{Ly$\alpha$}^\text{obs}$ are sufficiently widely scattered around the selection threshold, to examine if the selection effect can cause anisotropy in the IGM density similar to the observed one. 

Finally, by dropping objects whose $EW_\text{Ly$\alpha$}^\text{obs}$ is below some threshold, we obtain a sample of mock LAEs. We try two threshold $EW_\text{Ly$\alpha$}^\text{obs}$ values. One is $30$ {\AA} that is actually adopted to select the $z=2.2$ LAEs. The other is $40$ {\AA}; we try this slightly higher threshold to evaluate how sensitive the results are to the threshold. We select $35$ ($12$) O3Es with $EW_\text{Ly$\alpha$}^\text{obs}\geq30$ ($40$) {\AA} as mock LAEs.
Using those mock LAEs, we conduct the same analyses (CCF and 1D-CCF) as for the real galaxies.

\subsection{Results}
\label{sec:results_mockLAE}

Figure~\ref{fig:IM_mock} presents the results of the toy model.
A similar trend to that of real LAEs is seen in the $EW_\text{Ly$\alpha$}^\text{obs}\geq40$ \AA\ sample, whose CCF is lower on the far side up to $r=5~h^{-1}$~cMpc, implying a factor $1.5$ density difference within $r=3~h^{-1}$~cMpc with the FGPA. 
The 1D-CCF along the line-of-sight of this sample is also systematically lower on the far side over $|\Delta r_{\rm TRANS}|<6~h^{-1}$~cMpc (Figure~\ref{fig:IM_mock} (d)) in contrast to those along transverse directions, which are nearly symmetric over a similar distance (Figure~\ref{fig:IM_mock} (f)).
On the other hand, the $EW_\text{Ly$\alpha$}^\text{obs}\geq30$ {\AA} sample does not show significant line-of-sight anisotropy.
We note anisotropy of the 1D-CCFs along transverse directions seen beyond $|\Delta r_\text{TRANS}| \geq 5~h^{-1}$~cMpc in Figures~\ref{fig:IM_mock} (e) and (f). This could be due to the small sample size of mock LAEs because the anisotropic feature becomes weaker with increasing sample size, probably by mitigating the large-scale variation in the IGM {\sc Hi} distribution around them. 

Interestingly, the degree of anisotropy is sensitive to the threshold $EW_\text{Ly$\alpha$}^\text{obs}$ value.
The absence of anisotropy in the  $EW_\text{Ly$\alpha$}^\text{obs}\ge 30$ {\AA} sample may indicate that this \textit{EW} threshold is not high enough to produce detectable anisotropy in our mock LAEs.
We also find that the mock LAEs with 
$EW_\text{Ly$\alpha$}^\text{obs}\ge 40$ 
{\AA} show stronger anisotropy in the 1D-CCF along the line-of-sight than the real LAEs. It may be because O3Es are, on average, located in higher IGM density environments that give higher near-to-far density contrasts. 
Another intriguing fact is a smaller offset of the 1D-CCF along the line-of-sight between the near and far sides seen in Figure~\ref{fig:IM_mock} (d) than the variation of 
$\langle\delta_\text{F}\rangle^\text{LOS}_{near}$
among O3Es seen in Figure~\ref{fig:dist_EW}. It implies that this selection effect is detectable only statistically.

\subsection{
Possible impacts of Ly$\alpha$ profile shape and peculiar motion on LAE selection
}
\label{sec:discuss_mockLAE}

Here we briefly discuss the possible impacts of Ly$\alpha$ profile shape and peculiar motion on the selection of mock LAEs.
Changing the blue peak's width (currently 200 km s$^{-1}$) and position ($-200$ km s$^{-1}$) will select a different set of mock LAEs because of a change in the distance range over which Ly$\alpha$ photons are attenuated.
The IGM attenuation is sensitive to the fraction of the blue ($<\lambda$(Ly$\alpha$)) part ($24\%$ is assumed).
Galaxies with an expanding {\sc Hi} shell (e.g., \citealp{Verhamme08}) will have low blue fractions because of systematic redshifting of the entire Ly$\alpha$ profile.
As an extreme case, if galaxies have no blue part as found for some LAEs (e.g., \citealp{rakic11,hashimoto13,hashimoto15}), no IGM attenuation, and hence no selection effect, will occur. 
Galaxies with higher receding peculiar velocities than the IGM will suffer weaker selection effects because of decreased fractions of the Ly$\alpha$ emission subject to the IGM attenuation.
If the actual selection effect by the IGM attenuation is much weaker than assumed in our toy model, we have to invoke an alternative mechanism as the origin of the observed anisotropy.


\begin{figure*}
	\begin{center}
	\includegraphics[width=\linewidth]{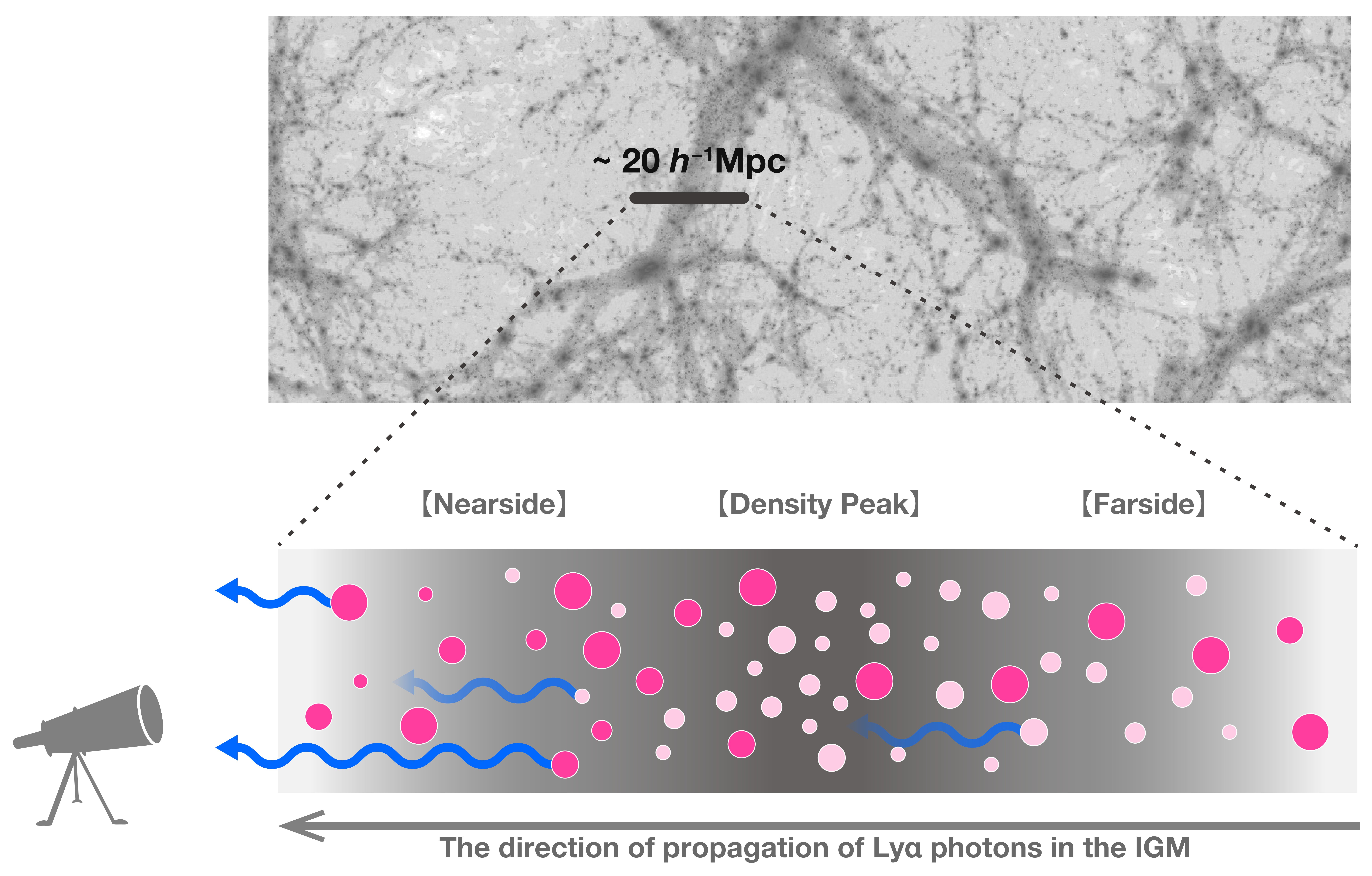} 
	\caption{Schematic picture of the selection bias against LAEs in a dense region. The IGM {\sc Hi} density is represented by the grayscale, with darker colors meaning higher densities. 
    Circles indicate galaxies, with larger sizes meaning higher intrinsic Ly$\alpha$ luminosities. The galaxy sample as a whole follows the matter/{\sc Hi} density distribution. It means that the number of galaxies is proportional to the matter/{\sc Hi} density where they are located; 
    the number of galaxies in the darkest area is highest. 
    The Ly$\alpha$ emission from each galaxy travels toward us from right to left in the IGM.
    Galaxies that are detectable in Ly$\alpha$ are colored in dark pink, while those not are in faint pink.
    Blue arrows indicate the direction and propagation of Ly$\alpha$ photons in the IGM.
    The Ly$\alpha$ photons emitted from a galaxy cannot reach us if the IGM {\sc Hi} in $3~h^{-1}$~cMpc on the near side of the galaxy exceeds a certain density (translucent arrows).
    }
	\label{fig:loci_gal}
	\end{center}
\end{figure*}


\section{Discussion and Summary}
\label{sec:summ}

We have investigated the IGM {\sc Hi} density traced by Ly$\alpha$ forest absorption around three galaxy populations at $z\sim 2$, particularly paying attention to LAEs. 
We have found that LAEs show an anisotropic IGM {\sc Hi} density distribution along the line-of-sight, while continuum-selected galaxies and O3Es do not.  
Such anisotropy is obtained if LAEs on the far side of dense regions tend not to be selected because of heavier Ly$\alpha$ attenuation by the IGM {\sc Hi}.
An examination with mock LAEs supports this idea, finding an anisotropic {\sc Hi} density distribution as observed.

Other mechanisms could also create an anisotropic IGM {\sc Hi} density distribution along the line-of-sight.
One candidate is feedbacks from galaxies. If some feedback energy of a galaxy is emitted toward us, {\sc Hi} gas in front of the galaxy may be swept out or ionized, thus decreasing the IGM {\sc Hi} density on the near side.
Quasars and Active Galactic Nuclei (AGN) jets can drive gas and/or energy in the host galaxies to the surrounding IGM up to distances of a few Mpc (e.g., \citealp{Perucho14,Dabhade17,Dabhade20,momose20b}). Moreover, \citet{Hopkins20} have indicated that cosmic ray (CR)-driven outflows extend beyond $\sim 1$ Mpc. 
However, none of our LAEs has an AGN signature \citep{naka12,konno16,momose20b}. 
CR-driven outflows are also unlikely because their effects on the surrounding IGM become negligible at $z\geq1-2$ \citep{Hopkins20}. 
Thus, feedbacks are unlikely to be the major cause of the anisotropic density distribution around our LAEs.

Taking all these results into account, we conclude that the observed anisotropy of the IGM {\sc Hi} distribution around LAEs is likely due to a selection effect of IGM absorption (Figure~\ref{fig:loci_gal}).
If a dense region is viewed from a different direction, some are missed from the original LAE sample, while some are newly selected as if galaxies are playing hide-and-seek using {\sc Hi} gas.
We note that anisotropy may also occur along with transverse directions because of complicated Ly$\alpha$ radiative transfer.
Nevertheless, its amplitude seems to be negligible, as we have presented in Sections~\ref{sec:result} and \ref{sec:results_mockLAE}.

Our results also indicate that LAEs may not faithfully trace matter distribution, unlike O3Es and continuum-selected galaxies.
Some observational studies have found a $3-15~h^{-1}$~cMpc offset between the overdensity peaks of LAEs and other galaxy populations \citep{toshikawa16,shimakawa17b,shi19}. 
Given the anisotropy of {\sc Hi} density distribution over $r=3-5~h^{-1}$~cMpc found in this study, the discordance reported by those previous studies can be, at least partly, explained by the selection bias. 
Hence, we should be careful when using LAEs to search for overdensities such as proto-clusters.

As has been found in Sections~\ref{sec:result} and \ref{sec:results_mockLAE}, an anisotropic {\sc Hi} density distribution seems to be detectable even when the {\sc Hi} density contrast of a given region is as small as a factor of $\sim2$.
Since such a small difference can be easily realized between the densest part and the outskirt of a cosmic filament, we expect an anisotropic distribution across $\sim10-20~h^{-1}$~cMpc can be observed if the sightline passes through a filament.

Our finding of an anisotropic {\sc Hi} distribution around LAEs is based on a small sample ($N=19$). To increase the statistical reliability, a larger sample from a wide-field spectroscopic survey, such as one planned with the Subaru Prime Focus Spectrograph (PFS), is necessary. 
Followup spectroscopy of Ly$\alpha$ emission of the sample galaxies is also useful.
Theoretically, there seems to be a debate on the impact of Ly$\alpha$ radiative transfer effects on galaxies in current simulations (e.g., \citealp{Zheng11a,Hough20}). Theoretical studies focusing on line-of-sight dependence of the IGM {\sc Hi}--galaxy connection over a wide range of cosmological environment are also needed.


\section*{Acknowledgements}

We appreciate an anonymous referee for useful comments to improve our manuscript. 
We are grateful to Dr. K.-G. Lee for providing the CLAMATO data. 
We thank Drs. H. Yajima and K. Kakiichi for helpful discussions. 
RM acknowledges a Japan Society for the Promotion of Science (JSPS) Fellowship at Japan. 
This work is supported by the JSPS KAKENHI Grant Numbers JP18J40088 (RM), JP19K03924 (KS), and JP17H01111, 19H05810 (KN). 
We acknowledge the Python programming language and its packages of numpy, matplotlib, scipy, and astropy \citep{astropy13}.
%


\appendix

\section{The density difference between the near and far sides}
\label{sec:app_FGPA}

The density difference between the near and far sides is evaluated with the Fluctuating Gunn--Peterson Approximation (FGPA, e.g., \citealp{rauch97,croft98a,Weinberg99,Becker15}). In the FGPA, optical depth, $\tau_\text{FGPA}$, is described by a power-law of normalized gas density, $\rho$/$\bar{\rho}$, with
\begin{equation}
    \tau_\text{FGPA} \propto \left( \frac{\rho}{\bar{\rho}} \right)^\beta .
\end{equation}
Then, the density difference is expressed as the ratio:
\begin{equation}
    \frac{\rho_\text{far}}{\rho_\text{near}} = \left( \frac{\tau_\text{FGPA}^\text{far}}{\tau_\text{FGPA}^\text{near}}  \right)^{1/\beta} .
\end{equation}
The $\tau_\text{FGPA}$ is calculated from $\langle\delta_\text{F}\rangle$ with:
\begin{align}
    \begin{split}
        \tau_\text{FGPA} 
        &= - \ln{\left( \frac{F}{\langle F_z \rangle} \right)} \\
        &= - \ln{(\delta_\text{F} + 1)} .  
    \end{split}
    \label{eq:tau}
\end{align}
In this study, we use $\beta=1.6$ (e.g., \citealp{croft98a,Weinberg99}) and calculate the {\sc Hi} density difference within $r=3~h^{-1}$~cMpc.

We estimate $\tau_\text{FGPA}^\text{far}$ and $\tau_\text{FGPA}^\text{near}$ for the 19 real LAEs, and obtain $\rho_\text{far}/\rho_\text{near}=2.1$.
Similarly, the mock LAEs satisfying the $EW_\text{Ly$\alpha$}^\text{obs}=30$ and $35$ {\AA} thresholds have $\rho_\text{far}/\rho_\text{near}=1.3$ and 2.1, respectively.

\bibliographystyle{aasjournal}
\bibliography{clam}

\end{document}